\documentclass{PoS}
\usepackage{amsmath}
\usepackage{graphicx}
\usepackage[small]{subfigure}
\usepackage{braket}
\usepackage{cancel}
\usepackage{xspace}
\usepackage{fmtcount}
\usepackage[small]{subfigure}
\usepackage{color}
\usepackage{multirow}

\newcommand{\df}{\mathrm{d}}
\newcommand{\Eop}{\hat{\mathcal{E}}_T}
\newcommand{\Oop}{\Omega_1(r)}
\newcommand{\ang}[1]{\tau_{(#1)}}

\title{Hadron Mass Effects in Power Corrections to Event Shapes\thanks{IFIC/13-01,
LPN13-009, MIT-CTP 4432}}

\ShortTitle{Hadron Mass Effects in Power Corrections to Event
Shapes}

\author{\speaker{Vicent Mateu} \\
        Center for Theoretical Physics, Massachusetts Institute of
          Technology, Cambridge, MA 02139, USA\\
          IFIC, UVEG - CSIC, Apartado de Correos 22085, E-46071, 
                  Valencia, Spain\\
        E-mail: \email{mateu@mit.edu}}

\author{Iain W.~Stewart\\
        Center for Theoretical Physics, Massachusetts Institute of
          Technology, Cambridge, MA 02139, USA\\
        E-mail: \email{iains@mit.edu}}

\author{Jesse Thaler\\
                Center for Theoretical Physics, Massachusetts Institute of
                  Technology, Cambridge, MA 02139, USA\\
                E-mail: \email{jthaler@mit.edu}}

\abstract{ %
We study the effect of hadron masses on the leading power correction of dijet
event-shape distributions. We define the transverse velocity operator, that
describes the effects of hadron masses. It depends on the ``transverse velocity''
$r$, which is different from one only for non-vanishing hadron masses.
We find that hadron-mass effects in general break
universality. However we provide a simple method to identify universality classes
of event shapes with a common power correction. We also compute the anomalous
dimension of the power correction and the structure of the corresponding Wilson
coefficient, finding a nontrivial result.
}

\FullConference{Xth Quark Confinement and the Hadron Spectrum,\\
		October 8-12, 2012\\
		TUM Campus Garching, Munich, Germany}

\begin{document}
\section{Introduction}
\label{sec:intro}\vspace*{-0.3cm}
Event shapes have been crucial to pin down the structure of QCD. In the recent years
the subject has attracted a lot of attention due to the very precise extraction of 
the strong coupling constant $\alpha_s$ from fits to the tail of the thrust distribution
\cite{Becher:2008cf,Abbate:2010xh,Gehrmann:2012sc} and moments of the thrust distribution
\cite{Gehrmann:2009eh,Abbate:2012}\footnote{See also \cite{Chien:2010kc} for a
determination using the Heavy Jet Mass distribution.}. Excellent reviews on event
shapes are \cite{Dasgupta:2003iq,Kluth:2006bw}, where the definition of the most
commonly used can be found.

Even though event shapes are infrared safe observables, they receive sizable corrections
from hadronization effects. In the tail of the distribution these effects are known as
power corrections and are suppressed by inverse powers of the center-of-mass energy
$Q$. The first studies of power corrections were inspired on renormalon techniques. The
dispersive approach of Dokshitzer and Webber \cite{Dokshitzer:1995zt,Dokshitzer:1995qm,
Dokshitzer:1998pt} replaces $\alpha_s$ by an effective coupling below some cutoff scale.
Within this approach it was found that the leading power correction for different event
shapes were proportional to one another, with a calculable coefficient
\cite{Dokshitzer:1995zt,Akhoury:1995sp}. Later, Salam and Wicke \cite{Salam:2001bd}
pointed out using the flux-tube model that hadron mass effects break that universality.

A different approach to power corrections is based on the factorization properties of
QCD at very high energies. The shape function introduced in \cite{Korchemsky:1999kt,
Korchemsky:2000kp} parametrizes nonperturbative corrections and describes the tail
power corrections to any order. Moreover, within this approach nonperturbative parameters
are expressed as matrix elements of QCD operators. Lee and Sterman \cite{Lee:2006nr}
showed that the factorization approach predicts the same universality relations as the
dispersive model. Recently it has been shown in the Soft-Collinear Effective Theory 
framework (SCET for short) \cite{Bauer:2000ew, Bauer:2000yr, Bauer:2001ct, Bauer:2001yt,
Bauer:2002nz} framework that hadron masses break universality \cite{Mateu:2012nk}.
%
\section{Event shapes in the dijet limit}
\vspace*{-0.3cm}
An event shape $e$ is an observable defined on the kinematical properties of the
final-state hadrons (energy and three-momentum). For our purposes we will use the
dimensionful quantities $p_\perp= |{\vec{p}_\perp}|$ (transverse momentum) and
$m^\perp = \sqrt{p_\perp^2+m^2}$ (transverse mass), and the dimensionless variables
$\eta=-\ln\tan(\theta/2)$ (pseudorapidity) and $y=1/2\ln[(E + p_z)/(E-p_z)]$ (rapidity).
Here $\theta$ refers to the polar angle from the $\hat z$ axis, which we take to be
aligned with the thrust axis. The velocity of a particle
is $v=|\vec p\,|/E$, and we define the transverse velocity as $r = p_\perp/m^\perp$.
For massless particles $p_\perp = m^\perp$, $y = \eta$ and $r = v = 1$, but the equalities
no longer hold for non-zero masses.

Dijet event shapes tend to zero for a configuration of two narrow back-to-back jets plus
soft radiation (that is, for a dijet configuration) and vice versa. In the dijet limit
$e\ll 1$ one can expand $e = {\bar e}+\mathcal{O}(e^2)$, where $\bar e$ is in general
simpler than $e$. In particular one can always write
\begin{align}\label{eq:f-definition}
\bar e = \frac{1}{Q}\sum_{i\in X} m^\perp_i f_e(r_i, y_i)\,,
\end{align}
where the function $f_e$ is specific for a given event shape. In this limit one can derive
factorization theorems for the differential cross section, which are very convenient to
perform resummation of singular logarithms to all orders in perturbation theory, and to
identify power corrections \cite{Bauer:2008dt}.
%
\section{Power Corrections}
\vspace*{-0.3cm}
Using SCET one can derive a factorization formula for the
singular cross section:
\begin{equation}\label{eq:fact-thmn}
 \frac{\df\sigma_s}{\df e}=\int \!\df \ell \:
   \frac{\df{\hat\sigma}_s}{\df e}\bigg(e-\frac{\ell}{Q}\bigg)\,F_e(\ell) 
  \, \big[ 1 +{\cal O}(e) \big] 
  \,.
\end{equation}
Here $\df{\hat\sigma}_s/\df e$ refers to the partonic singular distribution whereas
$\df\sigma_s/\df e$ is the nonperturbative singular distribution. $\df{\hat\sigma}_s/\df
e$ diverges as $\log^i(e)/e$ for $e\to0$ and hence dominates in the dijet limit. $F_e$ is
the shape function, which contains nonperturbative corrections (plus some perturbative
terms \cite{Mateu:2012nk}).

In the tail of the distribution, defined by the condition $Q\,e\gg \Lambda_{\rm QCD}$
the shape function can be expanded in inverse powers of $\ell \gg \Lambda_{\rm QCD}$:
\begin{align}\label{eq:shape-function}
F_e(\ell) &= \delta(\ell)  -\delta^\prime(\ell) \,\Omega_1^e(\mu) 
  +{\cal O}\Big(\frac{\alpha_s\Lambda_{\rm QCD}}{\ell^2}\Big) 
  + {\cal O}\Big(\frac{\Lambda_{\rm QCD}^2}{\ell^3}\Big) \, , 
\end{align}
Here $\Omega_1^e(\mu)$ is a nonperturbative matrix element defined by
\begin{align}\label{eq:Omega-definition}
\Omega^e_1 &= \langle\,0\,|\, \overline{Y}_{\bar n}^\dagger Y_n^\dagger
    (Q\hat e) Y_n \overline{Y}_{\bar n}\,|\,0\,\rangle\,.
\end{align}
and $\hat e$ is the event-shape operator defined as
\begin{align}
{\hat e}\,|\,X\,\rangle = e(X)\,|\,X\,\rangle\,,
\end{align}
with $\,|\,X\,\rangle$ the state of a configuration of particles in the final state of
a given event, and $e(X)$ the value of the event shape for that configuration. $Y$ and
$\bar Y$ are Wilson lines of soft gluon fields in the light-like directions $n$ and
$\bar n$.

Using Eq.~(\ref{eq:shape-function}) in (\ref{eq:fact-thmn}) one finds at leading order
that the effect of the power corrections is to shift the distribution:
\begin{equation} \label{eq:distnshift}
\frac{\df\sigma}{\df e}\, = \, \frac{\df \hat\sigma}{\df e} 
  - \frac{\Omega_1^e}{Q} 
 \frac{\df}{\df e} \frac{\df \hat\sigma}{\df e} +\ldots\,
 \, = \, \frac{\df \hat\sigma}{\df e}\bigg(e -\frac{\Omega_1^e}{Q}\bigg)
 +\ldots\,.
\end{equation}
A similar result is found in the dispersive model \cite{Dokshitzer:1995zt,
Dokshitzer:1995qm,Dokshitzer:1998pt}.
%
\section{Universality}
\vspace*{-0.3cm}
In order to study the effects of hadron masses on power corrections we need to express
the event-shape operator $\hat e$ in terms of quantum fields. Following the approach of
Refs.~\cite{Lee:2006nr,Bauer:2008dt} we find that it can be written in terms of the
energy-momentum tensor. Let us start by introducing the ``transverse velocity operator'',
defined by its action on a state $|\,X\,\rangle$:
\begin{align}\label{eq:trans-mom-flow-op}
 \Eop(r, y) \,|\,X\,\rangle 
  = \sum_{i \in X} m_i^\perp\delta(r-r_i)\, \delta(y - y_i) \,|\,X\,\rangle\,.
\end{align}
It is important to use rapidity $y$ and not pseudorapidity $\eta$ as only the former
transforms in an additive way under a longitudinal boost. In Ref.~\cite{Mateu:2012nk}
it was shown that $\Eop$ can be expressed solely in terms of the energy-momentum
tensor. Now the event-shape operator can be written as
\begin{align} \label{eq:e}
\hat{e} & \equiv \dfrac{1}{Q}\int_{-\infty}^{+\infty}\!\!\! \df y\,
  \int_0^1\!\! \df r \: f_e(r, y)\, \Eop(r, y) \,.
\end{align}
According the leading power correction is written in terms of a double integral:
\begin{align}\label{eq:jumptovacuum}
\Omega^e_1 &= \int_{-\infty}^{+\infty}\!\!\! \df y\,
  \int_0^1\!\! \df r \: f_e(r, y)
  \langle \,0 \,|\, \overline{Y}_{\bar n}^\dagger Y_n^\dagger
    {\cal E}_T(r, y) Y_n \overline{Y}_{\bar n} \,| \,0 \,\rangle\, .
\end{align}

\begin{figure}
\begin{center}
\includegraphics[width=0.5\columnwidth]{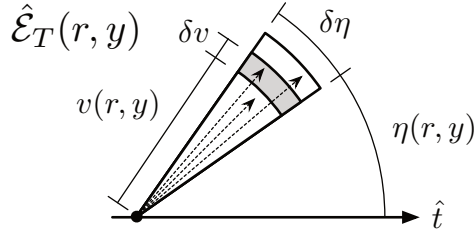}
\end{center}
\caption{Graphical representation of the transverse velocity operator.  Measurements are
made with respect to the thrust axis $\hat{t}$.  The arrows correspond to particles with
lengths given by the particle velocities. Shading indicates which particles are measured
by the operator.  Note that  the velocity $v(r,y)$ and pseudo-rapidity $\eta(r,y)$
are functions of the transverse velocity $r$ and rapidity $y$.}
\label{fig:transvelocityop}
\end{figure}

As depicted in Fig.~\ref{fig:transvelocityop}, the
transverse velocity operator $\Eop(r,y)$ involves a spheroid that expands in
both space and time with a finite velocity $v$, and it measures the total
transverse mass for particles in an infinitesimal interval in both $\eta$ and the
velocity $v$ (or equivalently an infinitesimal interval in $y$ and $r$).
Following Ref.~\cite{Lee:2006nr} one can apply boost transformations along the thrust
axis to figure out in which cases universality is preserved. Both the vacuum
$\,| \,0 \,\rangle$ and the Wilson lines are boost invariant, however under a boost of
rapidity $y^\prime$ the transverse velocity operator transforms as follows:
\begin{equation}\label{eq:boost-transformation}
U(y^\prime)\,\Eop(r, y)\,U(y^\prime)^\dagger= \Eop(r, y  + y^\prime)\,.
\end{equation}
Therefore, choosing $y^\prime = -y$ in Eq.~(\ref{eq:jumptovacuum}) we can write the
leading power correction as
\begin{align} \label{eq:O1univ}
  \Omega_1^e 
  &= c_e \int_0^1\!\! \df r \,g_e(r) \,\Omega_1(r) \,,
\end{align}
with
\begin{equation}
\label{eq:universalitya}
\Omega_1(r) \equiv \langle \,0 \,|\,  
   \overline{Y}_{\bar n}^\dagger Y_n^\dagger 
    \Eop(r, 0) Y_n \overline{Y}_{\bar n}
   \,| \,0\,\rangle
\end{equation}
a universal nonperturbative function, and
\begin{align} \label{eq:defncg}
&c_e = \int_{-\infty}^{+\infty}\!\!\!\!\! \df y \,f_e(1, y)\,,
&g_e(r) & = \frac{1}{c_e}\int_{-\infty}^{+\infty}\!\!\!\!\! \df y \,f_e(r, y)\,.
\end{align}
The function $g_e(r)$ encodes all hadron mass effects. Defining
\begin{align}\label{eq:universality}
\Omega^{g_e}_1 \equiv \int_0^1\!\! \df r \,g_e(r)\,\Oop\,,
\end{align}
one can write $\Omega^e_1 = c_e \,\Omega^{g_e}_1$, which implies that the leading power
correction for two event shapes $e_1$ and $e_2$ are proportional to each other if
$g_{e_1}(r) = g_{e_2}(r)$. We will denote the set of all event shapes with the same
$g_e(r)$ function as a universality class. All event shapes with the same universality
class have the same power correction up to a calculable factor.

The coefficients $c_e$ match exactly the classic universality prefactors obtained when
hadron masses are neglected \cite{Korchemsky:1994is, Dokshitzer:1995zt,
  Dokshitzer:1995qm, Dokshitzer:1997ew, Korchemsky:1999kt, Belitsky:2001ij,
  Berger:2003pk, Berger:2004xf, Lee:2006nr}. Table \ref{tab:cfunctions} summarizes the
values of the $c_e$ coefficients for the most common event shapes.
\begin{table*}[t!]
\begin{tabular}{l|ccccc}
 $\boldsymbol{c_e}$ 
  & \hspace{0.8cm}$\tau$\hspace{0.8cm} 
  & \hspace{0.8cm}$\tau_{2}$ \hspace{0.8cm} 
  & $\hspace{0.8cm}\tau_{(a)}\hspace{0.8cm}$ 
  & \hspace{0.8cm}$C$\hspace{0.8cm} 
  & \hspace*{0.8cm}$\rho_{\pm}\hspace*{0.8cm}$\\
\hline \\[-0.3cm]
Common~~ 
  & $2$
  & $2$ 
  & $\dfrac{2}{1-a}$ 
  & $3\pi$ 
  & $1$
  \\[3pt]
\end{tabular}
\caption{\label{tab:cfunctions}
Expression for the $c_e$ coefficients for various dijet event shapes.
Since $c_e$ are defined using $f_e(1,y)$, they have the same value in
each universality class. Here $\tau$ refers to thrust, $\tau_2$ to 2-Jettiness,
$\tau_a$ to angularities, $C$ to the C-parameter and $\rho_{\pm}$ to the hemisphere
masses.}
\end{table*}
%
\section{Mass schemes and universality classes}
\vspace*{-0.3cm}
The standard definition of an event shape involves in general both the energy and the
magnitude of the momentum of the final state hadrons (on top of the directions). In an
experimental environment one has access to a limited amount of information. Although
directions are easily measured, in general one has information on the energy deposited
by the particle in the detector, but not on its momentum. If the particle is identified
the momentum can be of course reconstructed, but that is not always possible.

The E-scheme is an alternative definition for any event shape in a way that only
the experimentally accessible information is used. Specifically one makes the following
replacement in the event-shape definition:
\begin{equation}\label{eq:E-scheme}
\vec{p}_i \to \frac{E_i}{|\vec{p}_i|}\,\vec{p}_i \,.
\end{equation}
It is easy to show \cite{Salam:2001bd,Mateu:2012nk} that all event shapes defined in the
E-scheme belong to the same universality class: the E-scheme class.

Analogously one can define the P-scheme class by the replacement $E_i \to |\vec{p}_i|$.
Although event shapes defined in that way do not belong to the same class, they have
nevertheless similar power corrections \cite{Salam:2001bd,Mateu:2012nk}.

We define two additional schemes: the R-scheme, in which one replaces $\eta$ by y in the
\mbox{P-scheme} expression of $\bar e$ and then uses $e^R = {\bar e}^R$; and the 
J-scheme, in which one sets $r = 1$ in the R-scheme expression.

Table \ref{tab:classes} summarizes which universality class event shapes (in various
schemes) belong to.
\begin{table*}[t!]
\begin{tabular}{l|c|l}
Class & ~~~$g(r)$~~~  & ~~Event shape\\\hline
Jet Mass class ($\Omega_1^0$ or $\Omega_1^\rho$) & $1$
  & ~~$\rho_\pm$, $\tau_{2}$, $\tau^J$, $\tau_{(a)}^J$, $C^J$ \phantom{\bigg(}
  \\
E-scheme class ($\Omega_1^1$ or $\Omega_1^E$)~~~ &  $r$
  & ~~$\tau_{(a)}$, $\tau^E=\tau_2^E$, $C^E$, $\rho_\pm^E$, $\tau^R$, $\tau_{(a)}^R$, $C^R$, $\rho_\pm^R$
  \\
$r^n$ class ($\Omega_1^n$)& $r^n$ 
& ~~generalized angularities $\ang{n,a}$ \cite{Mateu:2012nk} \phantom{\bigg(}\\
\hline
Thrust class ($\Omega_1^{g_\tau}$) & $g_\tau(r)$ 
& ~~$\tau$, $\rho_\pm^P$, $\tau_2^P$ \phantom{\bigg(}\\
$C$-parameter class ($\Omega_1^{g_C}$)~~& $g_C(r)$
& ~~$C$ \\
$r^2$ class ($\Omega_1^2$)  & $r^2$ & ~~$\ang{2,a}$, $\tau_{(a\to-\infty)}^P$\phantom{\bigg(}
\end{tabular}
\caption{Event shape classes with a universal first power correction parameter
$\Omega_1^{g_e}$. For a given event shape, the full power correction is
$\Omega_1^e = c_e \,\Omega_1^{g_e}$. Superscripts $E$, $P$, $J$, and $R$ correspond to
event shapes measured in the E-, P-, J-, and R-schemes, respectively.
\label{tab:classes}}
\end{table*}
\section{Anomalous Dimension and Matching Coefficient}
\begin{figure}[t!]
\begin{center}
\includegraphics[width=0.55\textwidth]{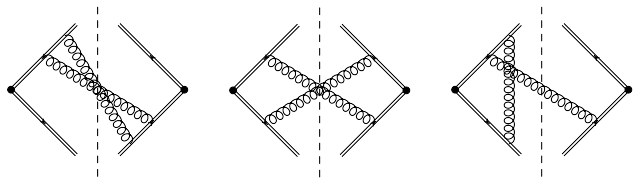}
\end{center}
\vspace{-0.5cm}
\caption{Independent emission diagrams with Abelian and non-Abelian
contributions.  The four additional diagrams obtained by a horizontal flip 
or complex conjugation are not shown. 
  \label{fig:non-abelian}}
\end{figure}

\begin{figure}[t!]
\begin{center}
\includegraphics[width=0.8\textwidth]{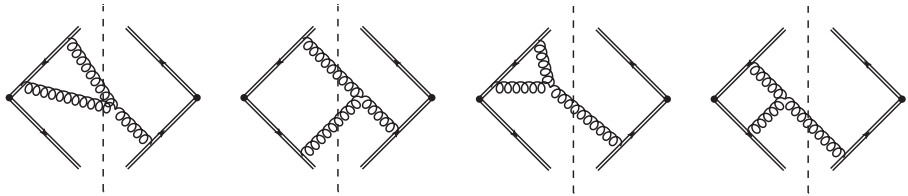}
\end{center}
\vspace{-0.5cm}
\caption{Triple gluon Y-diagrams for the ${\cal O}(\alpha_s^2)$ correction 
to $\Omega_1(r)$.
The twelve additional diagrams obtained by a horizontal flip 
or complex conjugation are not shown. Diagrams with all three gluons coupled to Wilson
lines of the same direction vanish.
\label{fig:triple-gluon}}
\end{figure}

\vspace*{-0.3cm}
The expression of $\Omega_1(r)$ in Eq.~(\ref{eq:universalitya}) is only a formal
definition. In general, matrix elements in a quantum field theory which do not directly 
correspond to an observable have to be defined within a scheme. In 
Ref.~\cite{Mateu:2012nk} the anomalous dimension of $\Omega_1(r)$ in the
$\overline{\rm MS}$ scheme was computed at one loop. The diagrams giving a non-vanishing 
contribution are shown in Figs.~\ref{fig:non-abelian} and \ref{fig:triple-gluon}. It 
turns out that  only non-abelian terms contribute, and one finds
\begin{equation}\label{eq:O1-anom-dim}
\mu\,\frac{\df }{\df \mu}\Omega_1(r,\mu) = \Big[-\frac{\alpha_s C_A}{\pi}\ln(1-r^2)
\Big] \ \Omega_1(r,\mu) \,.
\end{equation}
Interestingly, the anomalous dimension is $r$-dependent, although there is no mixing for
different values of $r$. This implies that hadron masses play an essential role.

From Eq.~(\ref{eq:O1-anom-dim}), for two renormalization scales $\mu$ and $\mu_0$ of
comparable size one has
\begin{align} \label{eq:expandedrunning}
\Omega_1^e(\mu) & = \Omega^e_1(\mu_0)+\frac{\alpha_s(\mu_0) C_A}{\pi}
\ln\Big(\frac{\mu}{\mu_0}\Big)\, \Omega^{e,\,\ln}_1(\mu_0)\,,
\end{align}
with
\begin{align} \label{eq:Oln}
\Omega^{e,\,\ln}_1(\mu_0) &\equiv - \!\int\! \df r \, \ln(1-r^2)\,
  c_e\,g_e(r)\,\Omega_1(r,\mu_0)\,.
\end{align}
Given that $\Omega_1(r,\mu)$ runs, one expects that the expansion of the shape function in
Eq.~(\ref{eq:shape-function}) should involve a non-trivial matching coefficient. Hence we write
\begin{align}\label{eq:shape-function2}
F_e(\ell) &= \delta(\ell) + \int \df r\,\, C_1^e(\ell,r,\mu)\:
  c_e\, g_e(r) \,\Omega_1(r,\mu) + {\cal O}\Big(\frac{\Lambda_{\rm QCD}^2}{\ell^3}\Big) \,.
\end{align}
Consistency with Eq.~(\ref{eq:O1-anom-dim}) requires that the matching coefficient at one loop
has the following form
\begin{align} \label{eq:C1e}
C_1^e(\ell,r,\mu) &= -\,\delta^{\,\prime}(\ell) 
  + \frac{C_A\alpha_s(\mu)}{\pi} \ln(1\!-\! r^2) \frac{\df}{\df\ell} \left(
   \frac{1}{\mu} \Big[\frac{\mu}{\ell}\Big]_+ \right)
   \nonumber\\
 &~~~
  + \frac{\alpha_s(\mu)}{\pi} \, \delta^{\,\prime}(\ell)\, d_1^e(r) 
  +{\cal O}(\alpha_s^2) \,.
\end{align}
The structure of Eq.~(\ref{eq:C1e}) was checked by an explicit calculation in
Ref.~\cite{Mateu:2012nk}.
\section{Conclusions}
\vspace*{-0.3cm}
We have studied hadron-mass effects for event shapes in the SCET formalism. These effects
have been expressed in terms of QCD matrix elements. Our results show that hadron masses
break universality although within certain classes it is still preserved.
We have computed the one-loop running of the power correction, finding a nontrivial
anomalous dimension and matching coefficient. We largely confirm the results of
Ref.~\cite{Salam:2001bd}.
\begin{acknowledgments}\vspace*{-0.2cm}
  This work was supported by the offices of Nuclear and Particle Physics of the
  U.S. Department of Energy (DOE) under grant numbers DE-FG02-94ER-40818 and
  DE-FG02-05ER-41360, the European Community's Marie-Curie Research Networks
  under contract PITN-GA-2010-264564 (LHCphenOnet) and by MISTI global seed funds.
  VM is supported by a Marie Curie Fellowship under contract PIOF-GA-2009-251174.
\end{acknowledgments}
\vspace*{-0.6cm}
\section*{}
\bibliography{thrust3}
\bibliographystyle{jhep}

\end{document}